\documentclass[dvipdfmx,prd,showpacs,preprint,preprintnumbers,amsmath,amssymb,eqsecnum,nofootinbib]{revtex4-1}
\usepackage{graphicx,amssymb,amsmath,amsthm,amsfonts,mathrsfs,epsfig,epsf}
\usepackage{dcolumn}
\usepackage{amsfonts,mathrsfs}
\usepackage{bm}
\usepackage{subfigure}
\usepackage{color}
\usepackage{url}
\usepackage{hyperref}
\hypersetup{
setpagesize=false,
 bookmarksnumbered=true,%
 bookmarksopen=true,%
 colorlinks=true,%
 linkcolor=blue,
 citecolor=red,
}

\usepackage[top=20truemm,bottom=20truemm,left=20truemm,right=20truemm]{geometry}

\newcommand{\D}{{\rm d}}
\newcommand{\eq}[1]{\mbox{Eq.~(\ref{#1})}}
\newcommand{\fig}[1]{\mbox{Fig.~\ref{#1}}}
\newcommand{\sign}{\text{sgn}}

\begin{document}


\title{Quasi-periodic relativistic shells in reflecting boundaries: \\
How likely are black holes to form?}

\author{Takafumi Kokubu} 
\email{takafumi.kokubu@ipmu.jp, 14ra002a@al.rikkyo.ac.jp}
\affiliation{Kavli Institute for the Physics and Mathematics of the Universe (WPI),
University of Tokyo, Kashiwa 277-8583, Japan}

\date{\today}          
\begin{abstract}
A system of two gravitating bodies floating around a restricted region of  strong gravitational field is investigated.
We consider two concentric spherically symmetric timelike shells spatially constrained by a perfectly reflecting inner and outer boundary. It is shown numerically that even when the gravitational radius of a contracting shell is larger than the radius of the inner boundary, energy transfer occurs due to the intersection with the other expanding shell before the contracting shell becomes a black hole, resulting nonlinearly stable motion.  The system appears to be in a permanently stable periodic motion due to the repetition of forward and reverse energy transfer. The larger the specific energy of a shell, the more stable the motion is. In addition, the motion of the null shell as the fastest limit of the timelike shell is also investigated. Unlike the timelike shell, the motion of the two null shells reduces to exact recurrence equations. By analyzing the recurrence equations, we find the null shells also allow stable motions. Using the algebraic computation of the recurrence equations, we show numerical integration is not necessary for the nonlinear dynamics of the null shells in confined geometry.
\end{abstract}
\maketitle
\section{Introduction}
Gravitational wave astronomy starting with the recently observed gravitational wave binary black hole merger showed that general relativity is still valid in strong gravitational fields \cite{LIGOScientific:2016aoc}.
Undoubtedly, the most attractive applications of general relativity are the phenomena associated with black holes (BHs).

The model of a gravitating body in motion due to self-gravity has been known for a long time \cite{Oppenheimer-Snyder}.
Research in recent decades has shown that when self-gravitating bodies or fields are spatially constrained, their behaviors become non-trivial. In confining geometries, it is highly non-trivial whether the final fate of the bodies/fields are a BH or stable periodic motion \cite{Bizon:2011gg, Buchel:2012uh, Maliborski:2012gx, Maliborski:2013jca, Okawa:2014nea, Okawa:2015xma}.
It is therefore important to investigate nonlinear gravitational phenomena in confined geometries to understand the nature of gravity.

We try in this paper to answer the question of how likely BHs are to form by considering the dynamics of spatially bounded gravitating bodies in a strong gravitational field.
In general, however, the nonlinear time evolutions of gravitational fields cannot be solved analytically due to gravitational wave emission from gravitational sources, and requires powerful numerical calculations.
 ``A gravitating shell'' successfully avoids such numerical difficulties.
 
 The notion of a shell, an infinitely thin matter-layer, is a simple and idealized gravitating matter. Israel formulated the full general relativistic behavior of the single shell incorporating self-gravity \cite{Israel1966}. Shell's nonlinear behavior have solved fundamental problems on nonlinear gravitational physics \cite{DrayHooft1985, redmount, PoissonIsrael1989}.
A few decades ago, an interesting method was developed for describing the nonlinear evolution of self-gravitating spherically symmetric concentric ``two shells'' \cite{Nunez+1993, NakaoIdaSugiura1999, IdaNakao1999, Barkov+2002JTEP}. This is a method to follow the time evolution after a crossing of the two shells, by assuming that each shell interacts only gravitationally at the crossing event. Hereafter we call this a two-shell system.

The motion of two gravitating shells, which can only gravitationally interact with each other, when spatially constrained, is described by a set of first-order ordinary differential equations and hence does not require powerful numerical methods for its numerical integration. Despite this simplification, it has been reported that this two-shell system capture  nonlinear gravitational phenomena, e.g., BH critical behavior, \cite{CardosoRocha2016, BritoCardosoRocha2016}. Thus, the two-shelf system is a good way to capture the nonlinear nature of gravity.
It is also known that spatially bounded two-shells generally exhibit chaotic properties \cite{Barkov+2002JTEP, MillerYoungkins1997, Barkov+2005, Kokubu:2020jvd}.

Now, to answer our question, we consider the following situation: Consider two concentric spherically symmetric dust shells spatially bounded by a perfectly reflecting rigid inner and outer boundary around a strong gravitational field. Assuming that the shells interact gravitationally, do they collapse into a BH or can they become nonlinearly stable?
It is obvious that if the gravitational radius of both shells is sufficiently smaller than the inner boundary, they will not become BH. On the other hand, it is non-trivial when the gravitational radius and the radius of the inner boundary are comparable. In fact, we will later show that the shells exhibit non-trivial behavior in the comparable situations.

We comment on the relevance of this study to ultra compact objects. The situation we consider is a strong gravitational field, where the inner reflective boundary is slightly larger than the Schwarzschild's gravitational radius $2M$ ($M$: gravitational mass). In other words, the inner boundary can be interpreted as horizonless ultra compact objects.
The gravastar is one of the horizonless ultra compact objects, which is connected to the inner de-Sitter spacetime and the outer Schwarzschild spacetime by a finite thickness of matter layers \cite{Mazur:2001fv}.
A model connected by an infinitesimally-thin shell has also been constructed \cite{Visser:2003ge}.
Since the gravastar is known to reflect almost all gravitational waves when the waves hit the gravastar's surface \cite{Pfister:1991ky, Pani:2009ss}, it may be possible to interpret our inner boundary as the surface of the gravastar. If such an interpretation is possible, our study can also be interpreted as a nonlinear analysis of the system consist of the massive shell and the gravastar.

The organization of the paper is as follows. 
In Sec. \ref{sec-setup}, we set up a shell model.
In Sec. \ref{sec-timelike-shell}, we find the motion of two timelike shells in our confined system by numerical integration. It is numerically observed that if the shells are moving very fast, the shells do not collapse, but make a quasi-periodic motion.
In Sec. \ref{sec-null-shell}, we consider the motion of two null shells, and show that  the expression for the shell's equation of motion is simplified by taking the light-speed limit of the timelike shell. 
We also show that the motion of the null shell is reduced to the analysis of an ``exact recurrence equations''. 
The null shell analysis gives us a helpful hint for the stable periodic motion observed in timelike shells.
Section \ref{sec-conclusion} is devoted to summary and discussion.

We take that the gravitational constant $G$ and the speed of light $c$ are unity.

\section{Setup}\label{sec-setup}
In this section, we setup crossings of two concentric dust thin-shells in the Schwarzschild spacetime. 
The line element of the Schwarzschild spacetime is given by
\begin{align}
\D s^2=-f(R)\D t^2+f(R)^{-1}\D R^2+R^2( \D \theta^2+\sin^2\theta\D \phi^2), \quad f(R)=1-\frac{2M}{R}.
\label{ds}
\end{align}
$M$ is the gravitational energy of the spacetime. 

\subsection{Single dust shell}
Let us first introduce a single shell as a timelike hypersurface which partitions the spacetime into the inner and the outer region.
On the hypersurface the line element is given by
$\D s_{\Sigma}^2=-\D \tau^2+r(\tau)^2 (\D \theta^2+\sin^2\theta\D \phi^2)$ with the shell's radius $r$ and the shell's proper time $\tau$.
By introducing Israel's junction conditions, Einstein equations for the shell are written as 
\begin{align}
8\pi S^a_{~b}=- [K^a_{~b}]+[K] \delta^a_{~b}, \label{hypersurface-eom}
\end{align}
where $K_{ab}$ is the extrinsic curvature and $S_{ab}$ is the stress-energy tensor of the dust,  $S^a_b={\rm diag}(-\rho, 0, 0)$ with the surface energy density $\rho$. The latin indice run over $\tau, \theta$ and $\phi$.
We defined a gap of a quantity $X$ at the hypersurface, $[X]:=(X_+-X_-)$. Subscript $+ ~ (-)$ denotes quantities in the outer (inner) region.
When the space time is given by \eq{ds}, the non-zero components of the extrinsic curvature are 
$K_{\tau \pm}^{\tau}=\dot \beta_\pm/\dot{r}$ and $K_{\theta \pm}^{\theta}=K_{\phi \pm}^{\phi}=\beta_\pm/r$, 
where $\dot{x} := \partial x/\partial \tau$ and $\beta_\pm:=\sqrt{f_\pm (r)+\dot{r}^2}.$
Junction condition \eq{hypersurface-eom} now reduces to
\begin{align}
-4\pi \rho&=(\beta_+- \beta_-)/r, \label{JUNCTION1} \\
0&=(\dot \beta_+-\dot \beta_-)/\dot{r}+(\beta_+- \beta_-)/r. \label{JUNCTION2}
\end{align}
Also, from the first fundamental form, one obtain
\begin{align}
\dot t_\pm:=\frac{\beta_\pm}{f_\pm(r)}.
\label{t-dot-root}
\end{align}
Because we consider a shell made of dust fluid, the following quantity is constant.
\begin{align}
m:=4\pi r^2 \rho.
\end{align}
$m$ denotes the shell's rest mass. We assume $m>0$ throughout this paper.
By squaring \eq{JUNCTION1}, the energy equation for the dust shell in the Schwarzschild spacetime with \eq{ds} is given by
\begin{align}
&\dot r^2+V(r)=0,  \quad
V(r)=1 -E^2-\frac{M_++M_-}{r} -\frac{m^2}{4r^2}
\label{dust-potential}
\end{align}
with the shell's specific energy $E$ and gravitational energy $\varepsilon$,
\begin{align}
\varepsilon:=M_+-M_-, \quad E:=\varepsilon/m.
\end{align}
Since $V(r\rightarrow \infty)=1 -E^2$, the shell marginally reaches $r=\infty$ with vanishing velocity for $E=1$ (marginally bound case). For $E>1$, the shell has a non vanishing velocity at infinity (unbound case). For $E<1$, the shell cannot reach infinity and bounded in finite radius (bound case).

\subsection{Two shells and crossings}
\begin{figure}[htbp]
  \begin{center}
    \begin{tabular}{c}
      \begin{minipage}{0.5\hsize}
        \begin{center}
          \includegraphics[scale=0.6]{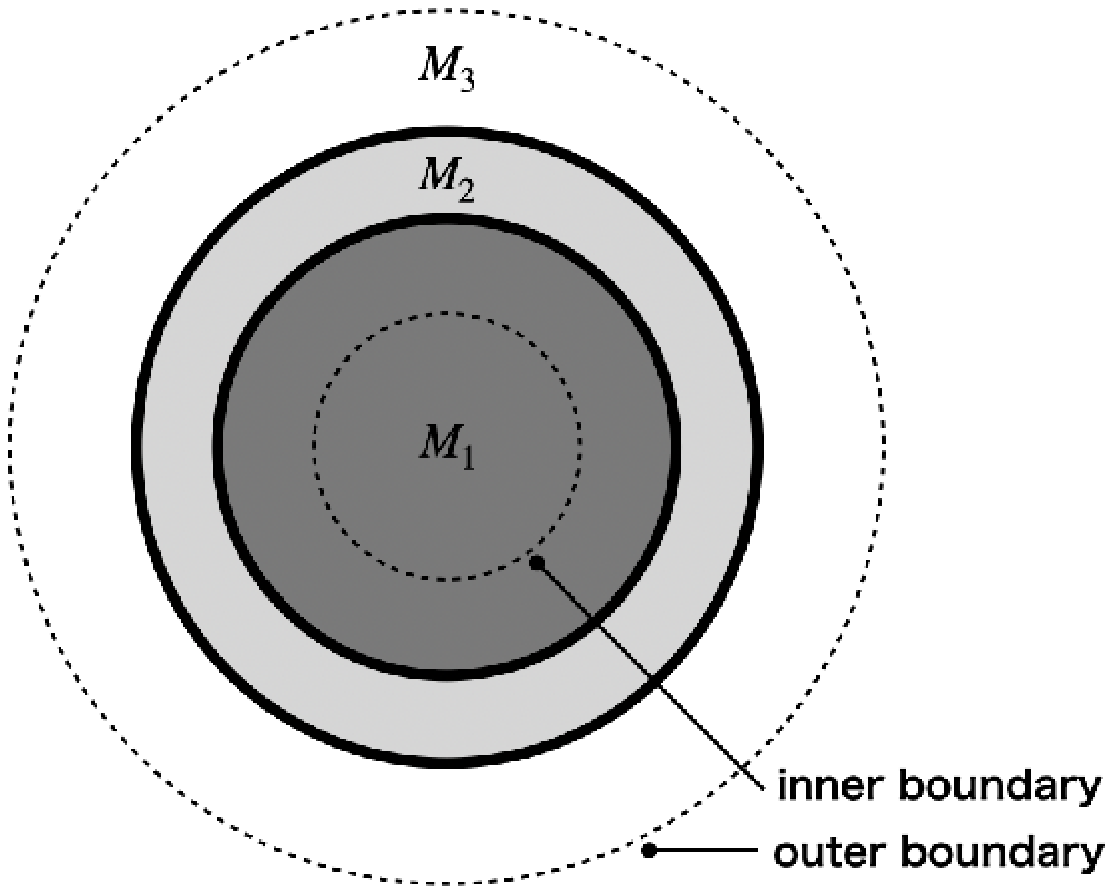}
          (a) 
        \end{center}
      \end{minipage}
      \begin{minipage}{0.5\hsize}
        \begin{center}
          \includegraphics[scale=0.9]{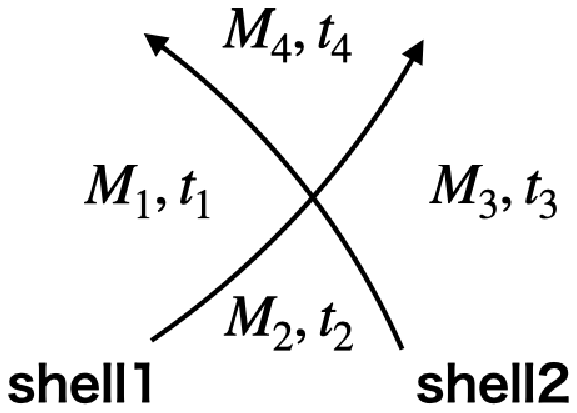}
         (b) 
        \end{center}
      \end{minipage}
\\
    \end{tabular}
    \caption{(a) Two shells (thick circles) between the inner and the outer boundaries (dashed circles), dark gray region with $M_1$, light gray region with $M_2$ and the outer white region with $M_3$.
    (b) Schematic picture of shell crossing. The inner shell (shell 1) and the outer shell (shell 2) cross at a crossing point.}    
\label{fig-collision}
  \end{center}
\end{figure}

To make a two shell system with confined geometry, we introduce the second dust shell outside of the first one. In addition, we place perfectly reflective inner and outer concentric boundaries in the spacetime. Shells are confined between the boundaries and will be purely reflected ($\dot r \rightarrow -\dot r$) when they hit the boundaries.
Thus, the presence of double shell inevitably divide the spacetime into four regions having distinct gravitational mass $M_I$ and time coordinates $t_I$ $(I=1,2,3,4)$.  \fig{fig-collision} explains our confined geometry and crossing of shells.
The equation of the inner shell with radius $r_1$ is obtained by taking $M_+=M_2$ and $M_-=M_1$ in \eq{dust-potential}, and similarly, the equation of the outer shell with radius $r_2$ is obtained by setting $M_+=M_3$ and $M_-=M_2$.  
We assume for simplicity the both shells have an equal rest mass, $m$.

To follow the time evolution of the shells after the crossing, at the crossing point we impose the transparent condition, i.e., the four-velocity and the rest mass of each shell are invariant during the crossing \cite{Nunez+1993, NakaoIdaSugiura1999}.
A possible change at the crossing is the gravitational energy $M_2$ in the region between shells. 
$M_2$ discontinuously varies at the crossing radius $r=R$ as $M_2 \rightarrow M_4$, where
\begin{align}
M_4=&M_3-M_2+M_1+\frac{1}{Rf_2}\left(M_2-M_1-\frac{m^2}{2R}\right)\left(M_3-M_2+\frac{m^2}{2R}\right) \nonumber \\
& -\sign \left(\frac{\D r_{1}}{\D \tau_{1}}\right) \sign \left(\frac{\D r_{2}}{\D \tau_{2}} \right)\frac{1}{Rf_2}
\sqrt{\left(M_2-M_1-\frac{m^2}{2R}\right)^2-m^2f_2}\sqrt{\left(M_3-M_2+\frac{m^2}{2R}\right)^2-m^2f_2}
\label{M2after}
\end{align}
with $f_2:=1-2M_2/R$. $sgn(x)$ is the sign function. The right hand side of \eq{M2after} is evaluated at $r=R$. 
After the crossing, \eq{dust-potential} is again applied to follow dynamics of each shell merely by replacing $M_2$ with $M_4$, but shell 1 (initially inner shell) becomes a new outer shell and shell 2 (initially outer) is a new inner shell.

To follow general relativistic multiple bodies, a common time coordinate must be adopted because the proper time for each body is in general different.
In the present case, to use a common time coordinate we take $t_2$, a time measured between shells. By multiplying \eq{dust-potential} by \eq{t-dot-root}, under the common time coordinate we have the following energy equation for the shell, labeled as $i ~(i=1, 2)$,
\begin{align}
\left(\frac{\D r_i}{\D t_2}\right)^2+\frac{f_2(r_i)^2V(r_i)}{f_2(r_i)-V(r_i)}=0.
\label{commontime-potential}
\end{align}

Let us discuss the energy of shells. Shells can transport energy to each other by crossing.
 If shell $i$ has gravitational energy $\varepsilon_i$, then the energy $\tilde \varepsilon_i$ changed by crossing can be written using the energy transfer $\Delta \varepsilon$ as follows \cite{ NakaoIdaSugiura1999}.
\begin{align}
\tilde \varepsilon_1=\varepsilon_1-\Delta \varepsilon \quad {\rm and} \quad
\tilde \varepsilon_2=\varepsilon_2+\Delta \varepsilon, 
 \label{energy-transfer}
\end{align}
where 
\begin{align}
\Delta \varepsilon :=\gamma m^2/R, \qquad 
\gamma:=-f_2\left(\frac{\D t_{1}}{\D \tau_{1}}\right)\left(\frac{\D t_{2}}{\D \tau_{2}}\right)  +f_2^{-1}\left(\frac{\D r_{1}}{\D \tau_{1}}\right) \left(\frac{\D r_{2}}{\D \tau_{2}}\right).
\label{delta-E}
\end{align}
$\gamma$ is evaluated at the crossing radius $r=R$ and denotes the Lorentz factor of the relative velocity between the shells. 
\eq{delta-E} means that $\Delta \varepsilon$ is always positive. This fact guarantees that inner shell always releases its energy to the outer shell.
We also note that from \eq{energy-transfer} there is energy conservation, 
\begin{align}
\tilde \varepsilon_1+\tilde \varepsilon_2=\varepsilon_1+\varepsilon_2.
 \label{energy-cons}
\end{align}

\section{Timelike shells: Numerical observations}\label{sec-timelike-shell}
In this section, we follow the time evolution of two timelike shells by numerically integrating \eq{commontime-potential}. 
As we can see from \eq{commontime-potential}, the dynamics of each shell is described by a first-order ODE.
Two shells cross each other many times because they are sandwiched by reflective boundaries, and the gravitational energy between the shells changes discontinuously at each cross. 
When enough energy is accumulated in one of the shells, that shell can collapse into a BH. 

We are interested in the motion of two shells, sandwiched between boundaries, starting from the same radius at the initial time in opposite directions to each other. Let the radii of the inner and outer boundaries be $r_{b1}$ and $r_{b2}$, respectively, and let $2M_1<r_{b1}<r_{b2}$. In particular, the radius of the inner boundary is taken as
\begin{align}
r_{b1}=(1+\epsilon)2M_1,
\end{align}
where $\epsilon$ is positive constant. When $1\gg \epsilon>0$, the inner boundary may be interpreted as the radius of the gravastar (See introduction).
We discuss here the relation between the gravitational radius of the shell and the boundary: When the gravitational radius of the inner and outer shell, given by $r_{g1}=2M_2$ and $r_{g2}=2M_3$, satisfies the relation $r_{g1}<r_{g2}<r_{b1}<r_{b2}$, the system is clearly  stable.
When $r_{b1}<r_{g1}<r_{g2}<r_{b2}$, the contracting shell (shell 1) immediately collapses into a BH. Therefore, nontrivial choice for possibly stable motions is limited to
\begin{align}
r_{g1}<r_{b1}<r_{g2}<r_{b2}. \label{initial-r}
\end{align}
Note that the relation of \eq{initial-r} means the gravitational radius of the outer shell is larger than the radius of the inner boundary.
In this paper, we investigate the motion starting from the initial relation of \eq{initial-r}.

There are still many free parameters left in this setup, so to simplify the discussion, we choose the parameters as follows.
\begin{align}
M_1=1,~ M_2=1+0.5\delta,~ M_3=1+\delta,~ m=\delta/A \quad (A=2,8,16). \label{initial-para}
\end{align}
$A$ determines $E$. When $A=2,8,16$, then $E=1, 2, 4$. The larger $E$, the greater the velocity at infinity and the closer to the speed of light. The $E=2,4$ correspond to an unbound case while $E=1$ is a marginally bound case.
Since $M_2$ changes at each crossing, $M_2$ in \eq{initial-para} represents the initial value.
From the above choice of $M_{1,2,3}$, the gravitational energy of each shell is the same at the initial time, i.e. $\varepsilon_1=M_3-M_2=\delta/2,\varepsilon_2=M_2-M_1=\delta/2$.

We also restrict the initial data to the following three types.
\begin{itemize}
      \item Type1 $(E=2): A=8, \epsilon=0.5, r_{b2}=20, r_0=7$.
      \item Type2 $(E=4): A=16, \epsilon=0.5, r_{b2}=20, r_0=7$.
      \item Type3 $(E=1): A=2, \epsilon=0.1, r_{b2}=3.5, r_0=2.5$.
\end{itemize}
$r_0$ is the shell's initial radius. It is of course possible to choose different initial data, but the above conditions are general enough. In other words, changing the initial conditions slightly different from those above does not qualitatively change the behavior of the motion.

\subsection{Type 1 initial data: $E=2$}
Under this initial data, the inner boundary is at $r_{b1}=3$.
For $\delta < 0.5$, BH formation is not possible in principle, thus the two-shell system is trivially stable. 
On the other hand, for $\delta \geq 0.5$, the gravitational radius of the outer shell is larger than or equal the radius of the inner boundary, yielding that the outer shell immediately forms a BH {\it if the inner shell is absent} (for $\delta \geq2.5$, the gravitational radius of the outer shell is larger than the initial radius). 

\begin{figure}[htbp]
  \begin{center}
          \includegraphics{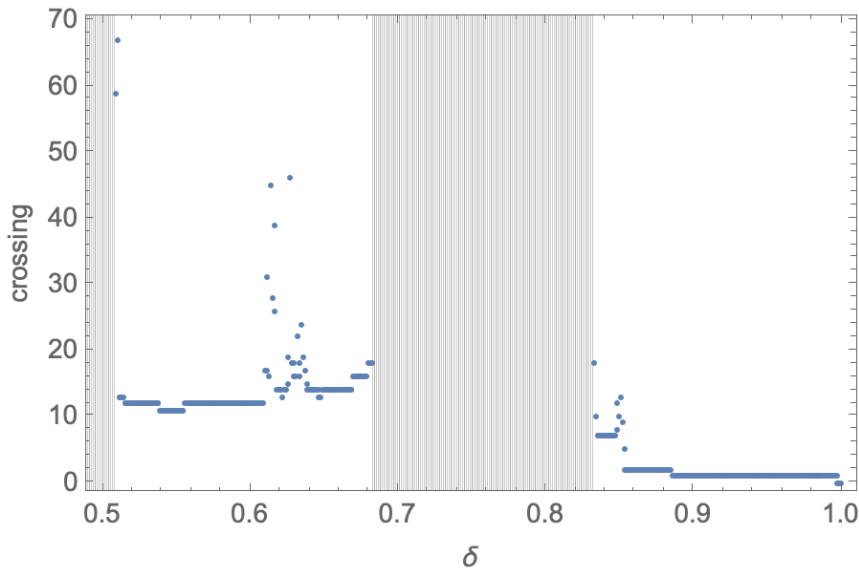}
    \caption{The number of crossings until BH formation, as a function of $\delta$. The gray areas in the figure correspond to stable solutions. $\delta < 0.5$ is trivially stable because BH cannot form in principle. Stable solutions exist in the non-trivial region of $0.683\lesssim \delta \lesssim0.832$.
        \label{fig-type1}}
  \end{center}
\end{figure}

Numerical integration of \eq{commontime-potential} with Type 1 initial data reveals the number of crossings until BH formation, as a function of $\delta$.
See \fig{fig-type1} \footnote{In the calculations of this paper we used {\tt Mathematica}; both {\tt AccuracyGoal} and {\tt PrecisionGoal} were chosen to be 12 digits of sufficient accuracy.}.
The gray areas in the figure correspond to stable solutions (stable means that the evolution does not collapse into a BH at least up to the integration time $t=2000$). We find nontrivial values of $\delta$ corresponding to the stable solution is
\begin{align}
0.683\lesssim \delta \lesssim0.832.
\end{align}
It is noteworthy that the stable region extends over a considerable range of $\delta>0.5$. This means that fine tuning is not necessary to obtain stable motion.
When $0.61 \lesssim \delta \lesssim 0.64$, one may notice that the number of crossings is higher than for other unstable solutions. In these deltas, the number of crossings changes sensitively with energy, but the evolution ultimately result in a BH.

\begin{figure}[htbp]
  \begin{center}
    \begin{tabular}{c}

      \begin{minipage}{0.5\hsize}
        \begin{center}
          \includegraphics[scale=0.5]{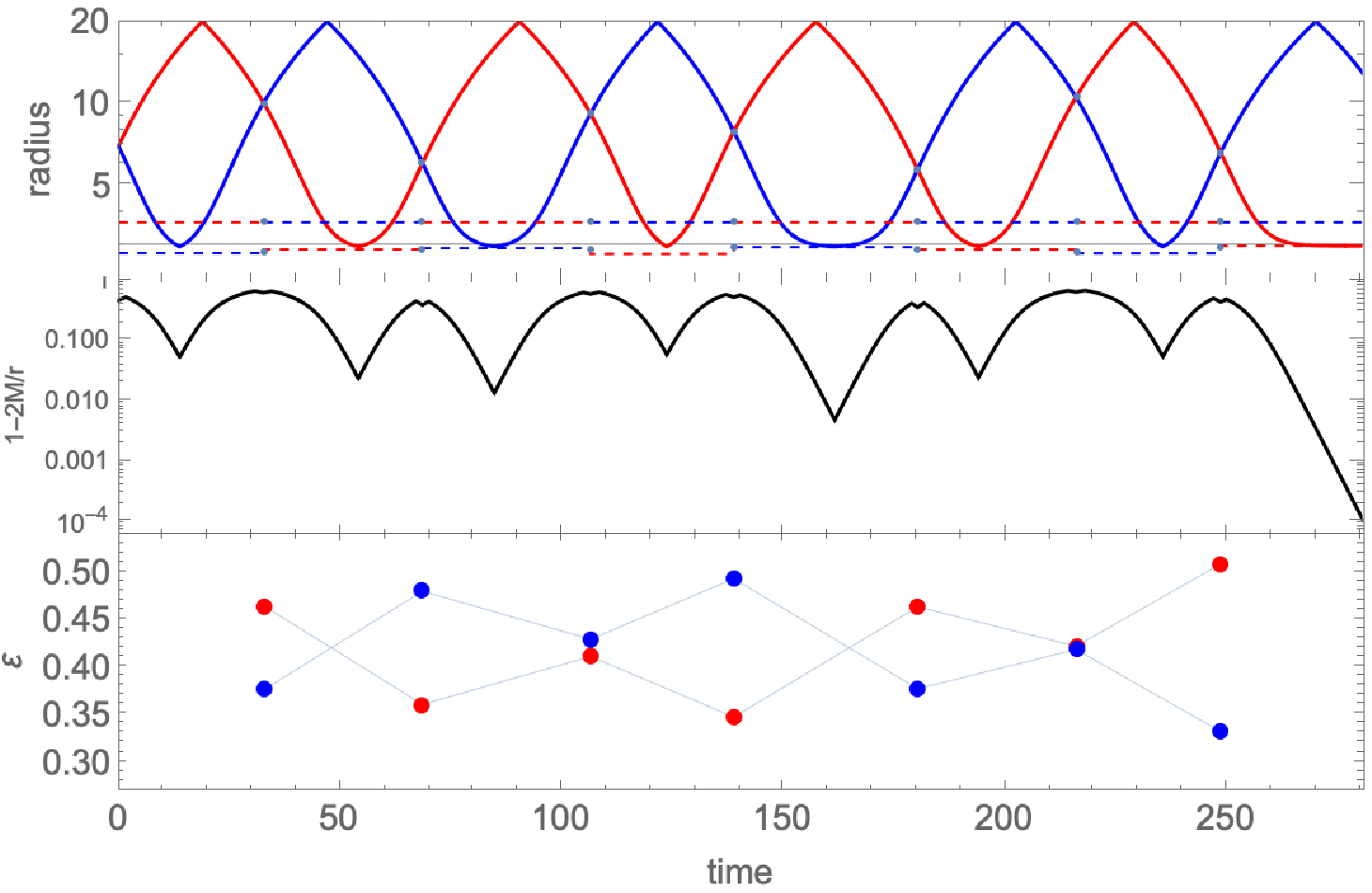}
          (a) 
        \end{center}
      \end{minipage}
      
      \begin{minipage}{0.5\hsize}
        \begin{center}
          \includegraphics[scale=0.5]{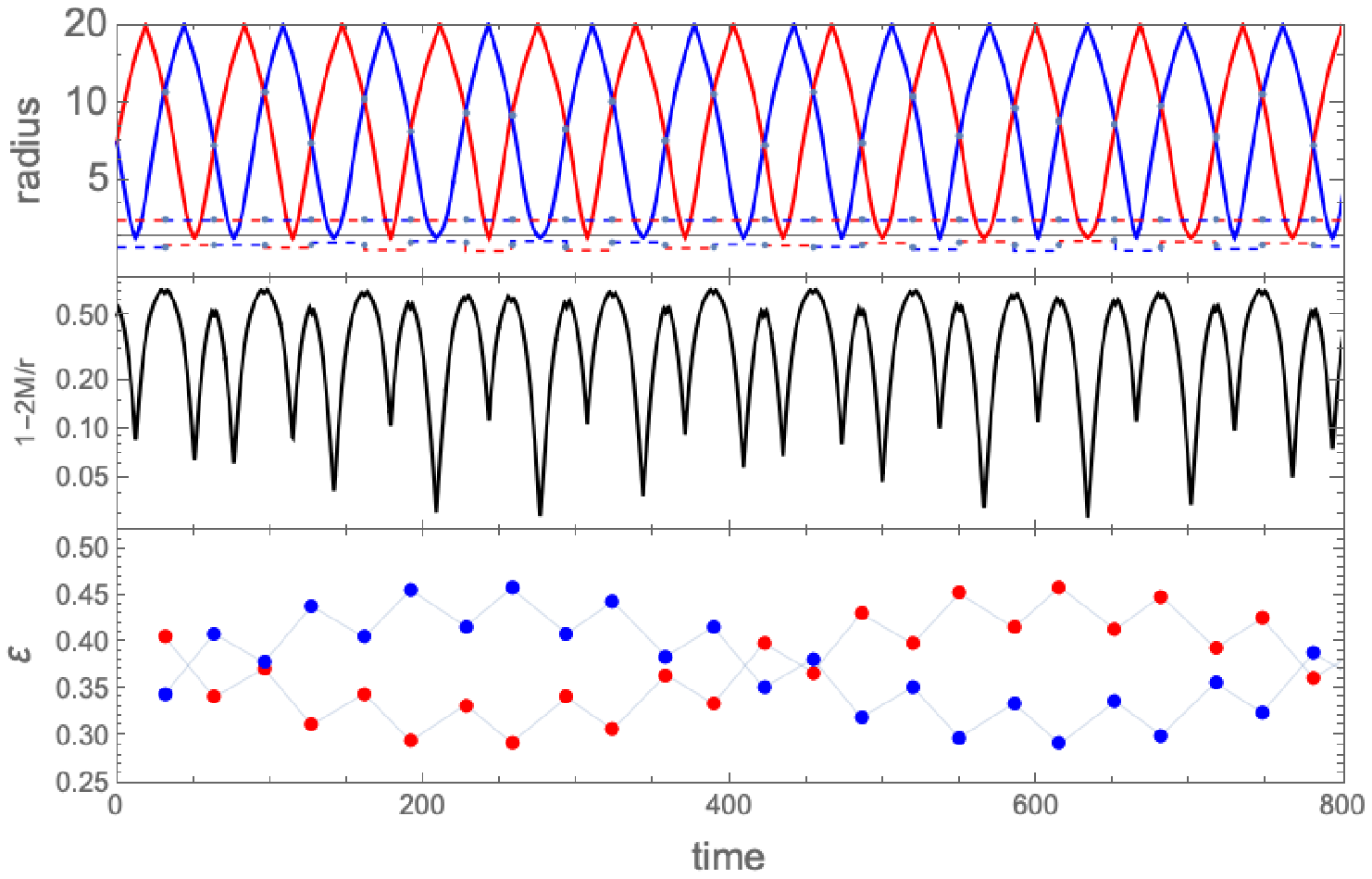}
         (b) 
        \end{center}
      \end{minipage}
      
\\
    \end{tabular}
    \caption{(a) {\bf Upper panel:} Time evolution of two timelike shells with $\delta=0.84$ for Type 1 initial data. After the seventh crossing, the red shell becomes a BH.
    {\bf Middle panel:} Lower value of $1-2M_2/r_1$ and $1-2M_3/r_2$. The zeroth of the value indicates a BH formation.
 {\bf Lower panel:} Shell's gravitational energy$\varepsilon$. The energy of the red shell eventually becomes larger than the energy of the blue shell and finally becomes a BH.
 (b) Stable quasi-periodic evolution with $\delta=0.75$ for Type 1 initial data. It behaves periodically by repeating energy forward and reverse transfer.}    
\label{fig-T1-1}
  \end{center}
\end{figure}

We show an example where the evolution eventually forms a BH. \fig{fig-T1-1}(a) represents the BH formation when $\delta=0.84$.
The upper panel represents the motion of each shell. Red/Blue shells are launched outwardly/inwardly. 
The black horizontal line is the radius of the inner boundary, and $r_{b1}=3$ for the Type 1 initial data.
The red dashed line represents the gravitational radius of the red shell and the blue dashed line is that of the blue shell. 
The position of these gravitational radii changes discontinuously at each crossing.
Middle panel is $\min\{1-2M_2/r_1, 1-2M_3/r_2\}$, and the point at which this value becomes zero represents the BH formation. After the seventh crossing, the red shell becomes a BH at $t_2\simeq 281$. 
Lower panel is the change in gravitational energy of the shell. since the total energy of the two shells is conserved (\eq{energy-cons}), each energy has a symmetric shape with respect to each other. As can be seen from the figure, the energy changes gradually with each crossing, and after the seventh crossing the red shell gains enough energy to become a BH.

Meanwhile, let us look at a stable solution with $\delta>0.5$. The solution for $\delta=0.75$ is \fig{fig-T1-1}(b). This is a ``stable quasi periodic motion''.
As can be seen from the upper panel of \fig{fig-T1-1}(b), the first crossing occurs before the outer shell (red shell) becomes a BH, and the decrease in the gravitational radius (dashed red lines) prevents the new inner shell (initially outer shell) from becoming a BH. Such a ``inner shell's sabotage'' continues throughout the whole evolution. 
Clearly, the presence of the second shell is preventing the first shell from becoming a BH.
The lower panel of \fig{fig-T1-1} (b) is the gravitational energy $\varepsilon$, and it is observed that the energy transfer is not one-way but periodic as a result of repeating forward- and reverse-transfer.

\begin{figure}[htbp]
  \begin{center}
    \begin{tabular}{c}

      \begin{minipage}{0.5\hsize}
        \begin{center}
          \includegraphics[scale=0.55]{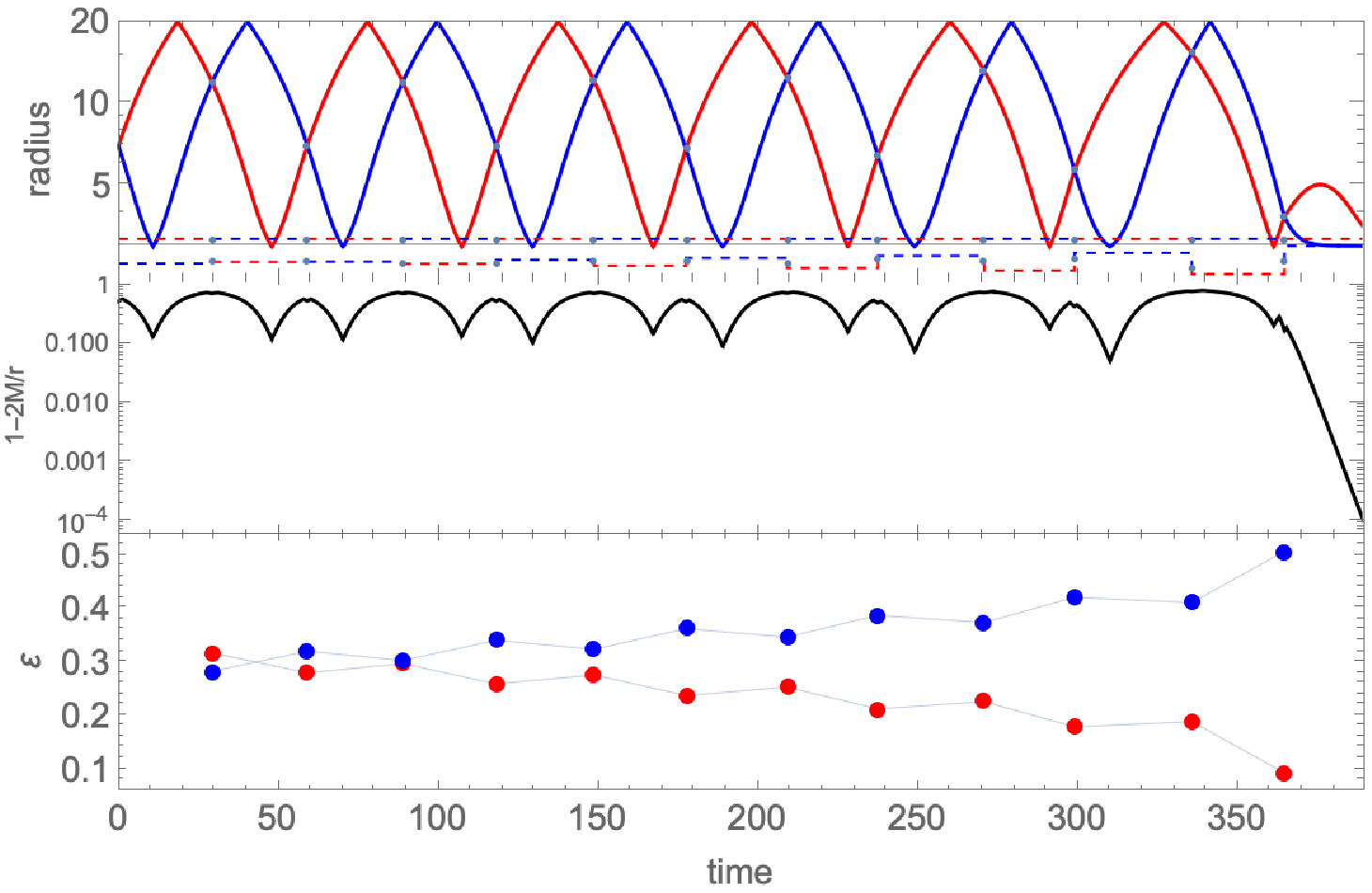}
          (a) 
        \end{center}
      \end{minipage}
      
      \begin{minipage}{0.5\hsize}
        \begin{center}
          \includegraphics[scale=0.55]{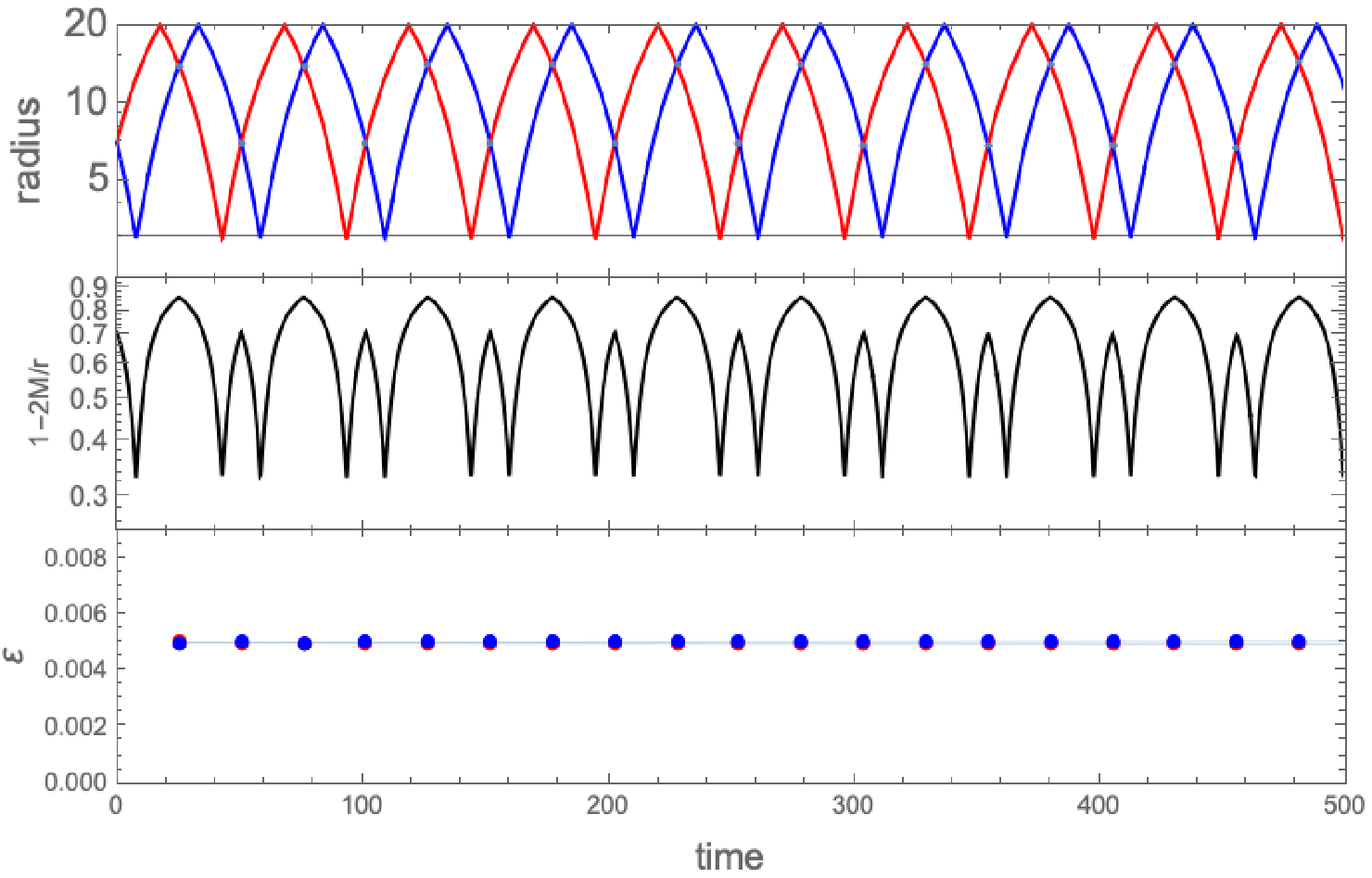}
         (b) 
        \end{center}
      \end{minipage}
 
\\
    \end{tabular}
    \caption{(a) BH formation with $\delta=0.6$ for Type 1 initial data.  As a result of multiple crossings, energy accumulates in the blue shell, and after the twelfth crossing, the blue shell finally collapses into a BH.
 (b) Stable quasi-periodic evolution with $\delta=0.01$ for Type 1 initial data. Two shells almost behave as test shells.}    
\label{fig-T1-2}
  \end{center}
\end{figure}

When $0.5 \lesssim\delta \lesssim 0.683$, BH formation again occurs. \fig{fig-T1-2}(a) is the BH formation when $\delta=0.6$. As a result of multiple crossings, energy accumulates in the blue shell, and after the twelfth crossing, the blue shell finally collapses into a BH.
The qualitative difference between BH formation in the large delta region  (i.e., $\delta \geq 0.832$) and the small delta region ($0.5 \lesssim\delta \lesssim 0.683$) is the following: In the small delta region, typically (but not always) one shell has so little energy that it cannot reach the outer boundary. This is because the total energy of the two shells is relatively small, so when energy is concentrated in the other shell due to energy transfer, the remaining shell will have less energy. \fig{fig-T1-2}(a) shows that this energy imbalance breaks the periodicity.
Also, although not shown in the figure, all motions with $\delta \lesssim 0.683$ always exhibit non-periodic or ``chaotic'' behavior (as far as we have confirmed numerically).

Now let us look at the motion when the delta is extremely small. \fig{fig-T1-2}(b) represents the motion with $\delta=0.01$. Here we also see a clear periodicity in the motion. This periodicity is due to the fact that the shell behaves almost as a test shell due to the small gravitational energy. Since there is almost no energy that can be transported, the shells have motions that are almost independent of each other. As a result, the crossing of the shells is restricted to only {\it two} specific radii (see appendix for the reason). Thus, the periodic motion with particularly small $\delta$ is qualitatively different from other periodic motions with larger $\delta$.

\subsection{Type 2 initial data: $E=4$}
\begin{figure}[htbp]
  \begin{center}
    \begin{tabular}{c}

      \begin{minipage}{0.5\hsize}
        \begin{center}
          \includegraphics[scale=0.55]{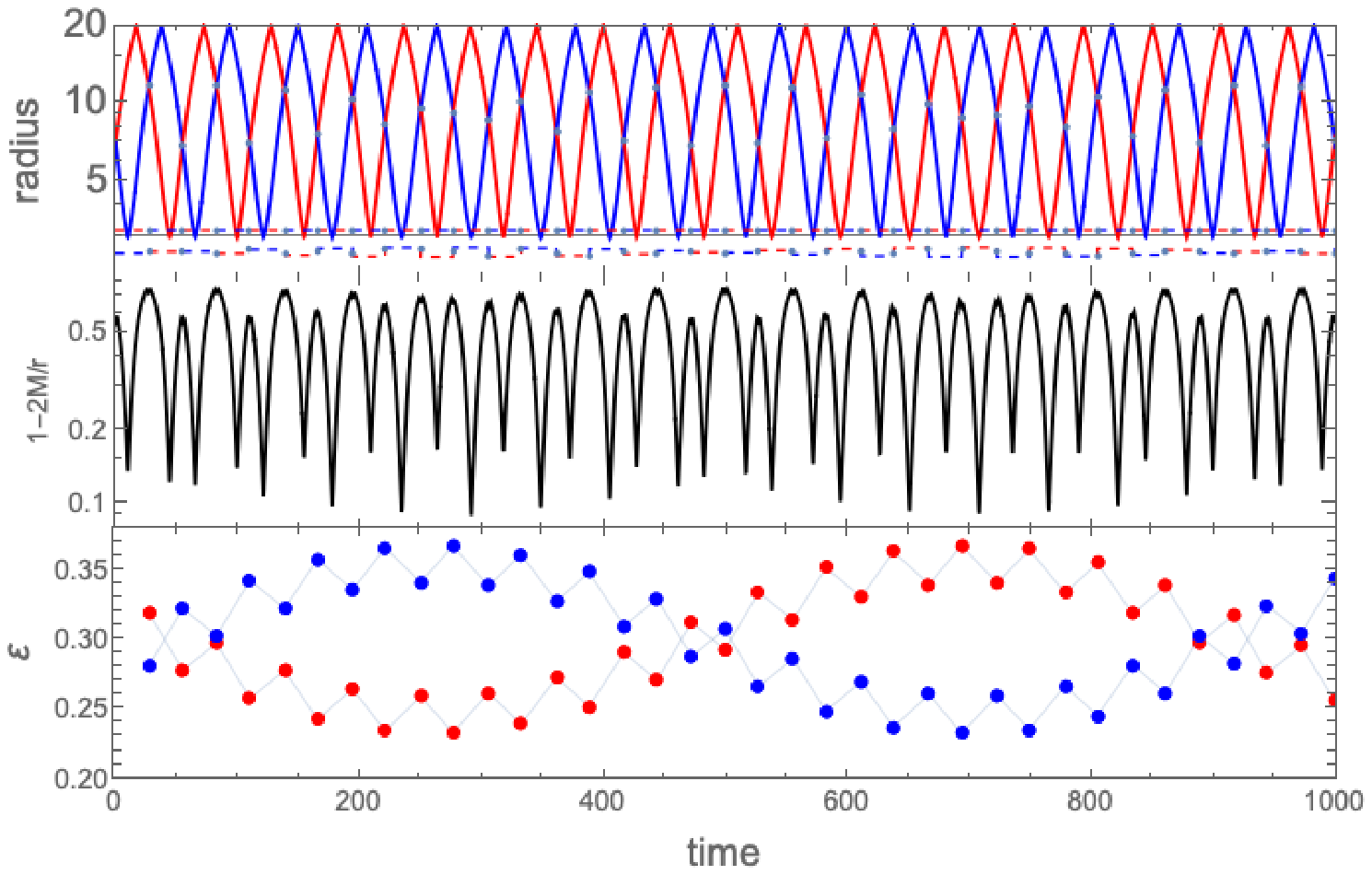}
          (a) 
        \end{center}
      \end{minipage}
      
      \begin{minipage}{0.5\hsize}
        \begin{center}
          \includegraphics[scale=0.55]{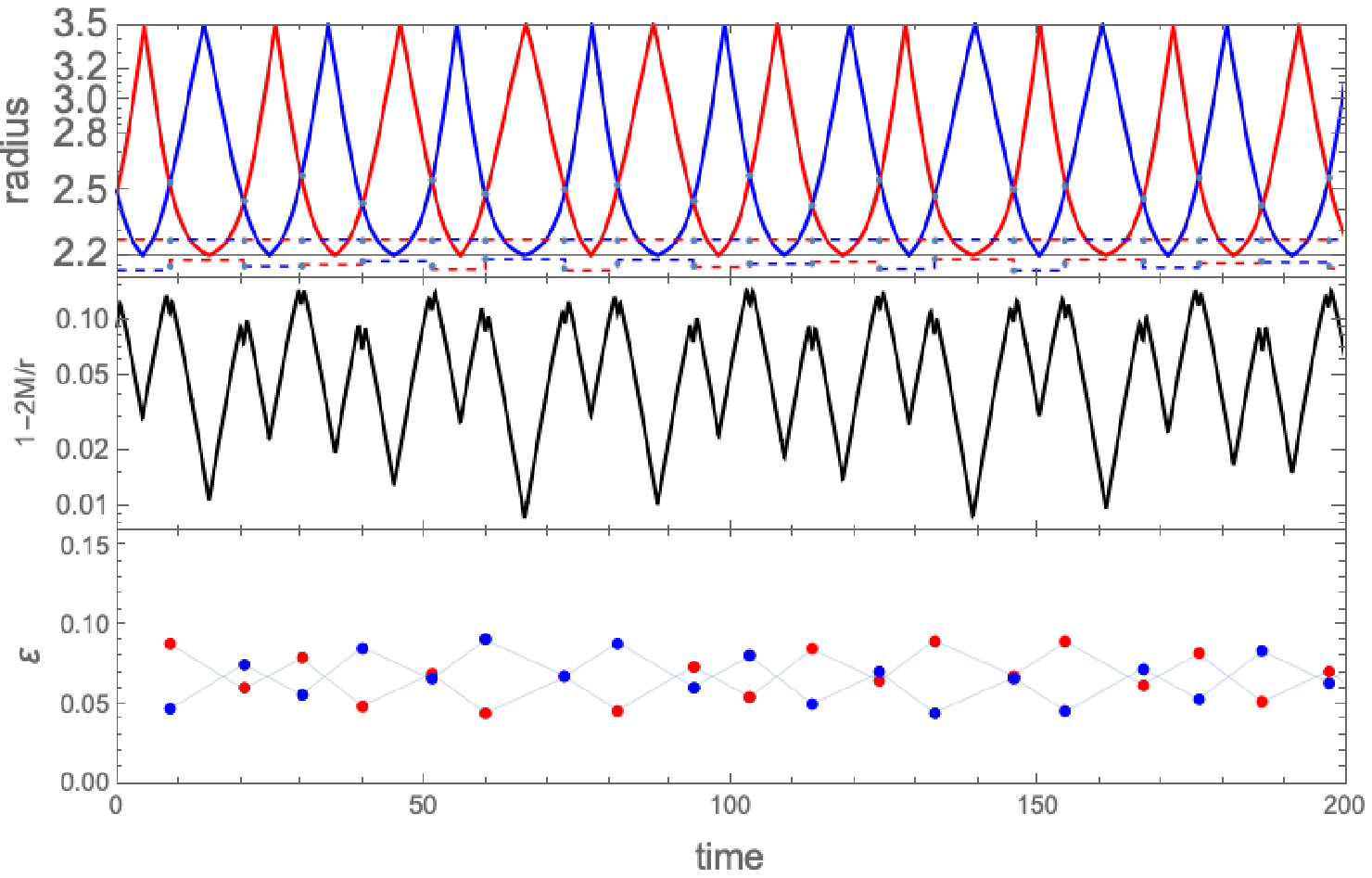}
         (b) 
        \end{center}
      \end{minipage}
 
\\
    \end{tabular}
    \caption{(a) Stable quasi-periodic evolution with $\delta=0.6$ for Type 2 initial data. (b)  Stable quasi-periodic evolution with $\delta=0.135$ for Type 3 initial data.}    
\label{fig-T2-T3}
  \end{center}
\end{figure}

In this type, the inner boundary is at $r_{b1}=3$.
There may exist nontrivial stable motion when $\delta \geq 0.5$. On the other hand, $\delta < 0.5$ does not allow BH in principle.
We numerically integrated the shell's equation of motion by continuously changing delta and found stable motion in the following range of $\delta$,
\begin{align}
\delta \lesssim 0.821.
\label{T2-stabledelta}
\end{align}
It can be seen that Type 2 has a wider distribution of stable regions than Type 1.
We plot the stable motion with $\delta=0.6$ in \fig{fig-T2-T3}(a). It can be seen that it is qualitatively the same as stable motions in Type 1.
Comparing Types 1 and 2, we see that the larger $E$ is, the more stable the region becomes. This implies that stable motion is more likely when the shell has relativistic speed. When $\delta$ is made extremely small ($\delta \ll 1$), the shell behaves like a test shell, as observed in Type 1.

\subsection{Type 3 initial data: $E=1$}
Type 3 is the initial data that restricts shells near the gravitational radius of the central body.
At this parameter, the inner boundary is at $r_{b1}=2.2$. For $\delta < 0.1$, it is trivially stable, i.e., the gravitational radius of the outer shell is smaller than the inner boundary, thus BH is not possible in principle. Therefore $\delta \geq 0.1$ corresponds to the possibility of nontrivial motion.
So far, we have focused our observations only on the motion of shells with large specific energy $E>1$. These shells are sufficiently relativistic at infinity. On the other hand, Type 3 is the marginally bound shells with zero velocity at infinity, denoting shell are energetically weak. 
However, even this ``weak'' shell can be relativistic near the gravitational radius of the central body because shell's kinetic energy becomes large due to strong gravity of the central body.
Numerical integration with continuously varying delta shows that stable motion exists, albeit within a relatively narrow range,
\begin{align}
0.132 \lesssim \delta \lesssim 0.138.
\end{align}
We show \fig{fig-T2-T3} (b), the time evolution of shells with $\delta=0.135$ for Type 3.
When the delta is extremely small ($\delta \ll 1$), it behaves like a test shell, similar to other types of initial data.

\section{Null shells: Exact algebraic analysis}\label{sec-null-shell}
In this section we analyze null shells as the light-speed limit of timelike shells. Null shells can be interpreted as a rough approximation of  gravitational shock waves \cite{DrayHooft1985}.
As can be seen from the discussion in the previous section, the evolution of two timelike shells can be thought of as a ``problem of finding the crossing points of two waves whose frequencies discontinuously vary with each crossing''. It is generally difficult (or simply impossible) to solve this problem in analytic sense.
In this section we consider the most relativistic situation where the motion is accelerated to the speed of light. 
As we will see below, the analysis becomes surprisingly simple in this extremal situation. More specifically, the motion of null shells can be integrated in the $(t_2, r)$ coordinate, and furthermore, the motion of the two shells between the boundaries is reduced to ``exact simultaneous recurrence equations''.
This means that numerical integration of equation of motions is not necessary to analyze the nonlinear dynamics of the null shell under consideration. 
We will discuss periodic motion and BH formation by analyzing the recurrence equations.
It will be explained how the analysis of null shells can help us understand the stable periodic motion of timelike shells.

\subsection{Exact recurrence equations}
The energy equation for the null shell is obtained by taking the light-speed limit of the timelike equation \eq{dust-potential}. For this porpose, we formally take the limit $m\to 0$ in \eq{dust-potential}. This limit diverges in the expression of proper time, but in the time coordinate between shells ($t_2$), the potential $\tilde V$ is finite. Formally taking the null limit for timelike shells of \eq{commontime-potential}, the potential takes the following form \cite{IdaNakao1999},
\begin{align}
\tilde V_{null}:=\lim_{m \rightarrow 0}\tilde V=-f_2(r)^2.
\end{align}
Thus, the energy equation becomes
\begin{align}
\left(\frac{\D r_i}{\D t_2}\right)^2+\tilde V_{null}(r_i)=0
\qquad (i=1,2).
\end{align}
This equation can be easily integrated with respect to $r_i$. The trajectory of shell $i$ starting from the initial radius $r_0$ is given by
\begin{align}
t_2=\sigma_i \left(r_i-R_0+2M_2 \log\left|\frac{r_i-2M_2}{R_0-2M_2}\right|  \right), \label{null-t-r}
\end{align}
where $\sigma_i$ takes $\pm1$, denoting an expanding shell for $+1$ and a contracting shell for $-1$.
Now, what we are interested in is the motion of two shells, sandwiched between two boundaries, starting from the same radius  in opposite directions. The inner/outer shells going in opposite directions reflect off the inner/outer boundary, and change its direction of motion, eventually approaching each other. After that, the two shells cross at a certain time.
Let us find the crossing radius, say, $R_1$. Since shell 1 moves inward and shell 2 moves outward, $\sigma_1=-1, \sigma_2=+1$ at the initial time. 
$\sigma_i$ changes its sign each time a shell reflects off a boundary. We assume that the shells do not form a BH by the time they crosses. We also assume that $2M_2<r_{b1}\leq r_1\leq r_2\leq r_{b2}$.
Shell 1 departing inwardly reaches the crossing radius $R_1$ after being reflected by the inner boundary $r_{b1}$. Let $T$ denote the time taken for this process.
Similarly, shell 2 departing outwardly reaches $R_1$ after being reflected by the outer boundary $r_{b2}$. Since the time taken for this process is also $T$, the following equation follows from \eq{null-t-r}.
\begin{align}
R_1-R_0-4r_{b1}&+2M_2 \log\left|\frac{(R_1-2M_2)(R_0-2M_2)}{(r_{b1}-2M_2)^2}\right| \nonumber \\
&=-R_1-R_0+4r_{b2}-2M_2 \log\left|\frac{(R_1-2M_2)(R_0-2M_2)}{(r_{b2}-2M_2)^2}\right|.
\end{align}
Solving this equation for $R_1$ yields
\begin{align}
R_1=2M_2\left(1+W\left( \frac{x_{b1}e^{x_{b1}}x_{b2}e^{x_{b2}}}{x_0e^{x_0}} \right)\right), 
\quad x_{b1,b2}:=\frac{r_{b1,b2}}{2M_2}-1,
\quad x_0:=\frac{R_0}{2M_2}-1,
\end{align}
where $W(x)$ is the prinpal branch of Lambert's $W$ function \cite{Corless:1996zz}.
Since we are assuming periodic motion here, there is also a second crossing, which is the intersection of the two shells in opposite direction starting at $R_1$. 
Obviously, the above calculation can be repeated. It is important to note that the gravitational energy between the shells is no longer $M_2$ after the first cross; the first crossing has changed the value of this energy. If we take the null limit, the new energy value, $\tilde M_2$, is determined by the following relation \cite{IdaNakao1999}.
\begin{align}
\left(1-\frac{2\tilde M_2}{R_1}\right)\left(1-\frac{2 M_2}{R_1}\right)
=\left(1-\frac{2M_1}{R_1}\right)\left(1-\frac{2M_3}{R_1}\right).
\end{align}
This relation is well known the Dray-'t Hooft-Redmount relation (DTR relation). See Refs.\cite{DrayHooft1985, redmount} for detailed discussions on the DTR relation.

Obviously, the same procedure can be repeated to find the second and subsequent crossing radius. Eventually, the $(n+1)$th crossing radius $R_{(n+1)}$ is obtained using the $n$th crossing radius $R_{(n)}$ and the gravitational energy $M_{2(n)}$ as follows.
\begin{align}
R_{(n+1)}=2M_{2 (n)}\left(1+W\left( \frac{y_{b1}e^{y_{b1}}y_{b2}e^{y_{b2}}}{y_n e^{y_n}} \right)\right), 
\quad y_{b1,b2}:=\frac{r_{b1,b2}}{2M_{2 (n)}}-1,
\quad y_n:=\frac{R_{(n)}}{2M_{2 (n)}}-1,
\label{Rn}
\end{align}
where $n=0, 1, 2, 3, \dots$ and $R_{(0)}, M_{2 (0)}$ denote the radius and the gravitational energy between the shells at the initial time.
Also, the $(n+1)$th gravitational energy $M_{2 (n+1)}$ can be easily deduced from the DTR relation as
\begin{align}
\left(1-\frac{2 M_{2(n+1)}}{R_{(n+1)}}\right)\left(1-\frac{2 M_{2(n)}}{R_{(n+1)}}\right)
=\left(1-\frac{2 M_1}{R_{(n+1)}}\right)\left(1-\frac{2M_3}{R_{(n+1)}}\right).
\end{align}
Or equivalently, solving the above relation with respect to $M_{2 (n+1)}$, we have
\begin{align}
M_{2 (n+1)}=\frac{(M_1+M_3-M_{2(n)})R_{(n+1)}-2M_1M_3}{R_{(n+1)}-2M_{2(n)}}.
\label{DTR}
\end{align}
The set of \eq{Rn} and \eq{DTR} is a simultaneous recurrence equations for two variables $R_{(n)}$ and $M_{2 (n)}$.
Given initial values $R_{(0)}$ and $M_{2 (0)}$, we can {\it algorithmically} obtain $R_{(n)}$ and $M_{2 (n)}$ at any $n$ by \eq{Rn} and \eq{DTR} {\it without directly performing numerical integration of equations of motion}. Thus, the dynamics of null shells that cross multiple times is reduced to the problem of dealing with the simultaneous recurrence equations.

Let us now discuss the conditions of BH formation.
The condition for the formation of a BH after the $n$th crossing is that the inner shell or outer shell contracts to its gravitational radius. In terms of the recursion relation, this BH criterion reduces to
\begin{align}
2M_{2(n)} \geq r_{b1} \quad {\rm or} \quad  
R_{(n+1)}\leq2M_3.
\label{BH-criterion}
\end{align}

\subsection{Evolutions of confined null shells}
We solve the time evolution of the null shells using the recursion relations \eq{Rn} and \eq{DTR}.
Although we have obtained the exact recursion relations, it still seems to be difficult to obtain the general term because of the nonlinearity of the recursion relations. 
However, we can demonstrate iteration of the recursion relations to check whether a given initial data result in BH formation  or periodic motion.
When a shell forms a BH during its evolution, $R_{(n)}$ or $M_{2(n)}$ meets the BH criterion \eq{BH-criterion} while increasing $n$. 
On the other hand, when the solution is not a BH but an oscillating solution, $R_{(n)}$ or $M_{2(n)}$ does not meet the BH criterion by increasing $n$. 

\begin{figure}[htbp]
  \begin{center}
          \includegraphics[scale=0.6]{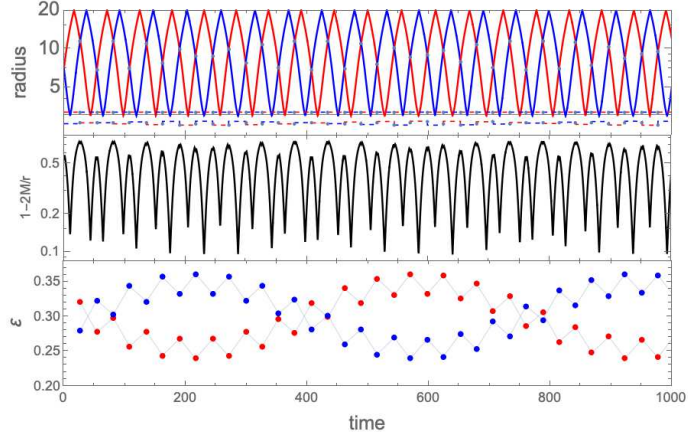}
    \caption{Stable quasi-periodic evolution of two null shells with $\delta=0.6$.
              \label{fig-NULL-STABLE06}}
  \end{center}
\end{figure}

To compare with the evolution of the timelike shells, we consider the following initial conditions.
\begin{itemize}
      \item $\epsilon=0.5, r_{b2}=20, r_0=7$
\end{itemize}
With these initial parameters the inner boundary is at $r_{b1}=3$. 
For $\delta < 0.5$, BH is not possible in principle. On the other hand, for $\delta \geq 0.5$, there are two possibilities of BH or stable motion.
After continuously varying $\delta$ and iterating the recurrence equations \eq{Rn} and \eq{DTR} up to $n=1000$, we found that the evolution exhibits stable periodic motion in the following range.
\begin{align}
\delta \lesssim 0.817.
\end{align}
It is noteworthy that this value is quite close to the stable region of the timelike shell in Type 2 (\eq{T2-stabledelta}). In fact, shells with Type 2 is quite fast since it is a motion with $E=4$.
The stable evolution of the null shell with $\delta=0.6$ is shown in \fig{fig-NULL-STABLE06} (here, we did numerical integration of equation of motion just to compare with the stable motion of the timelike shell). 
As can be seen from this figure, it is very similar to the timelike shell motion with the same delta in Type 2 (\fig{fig-T2-T3} (a)).

\begin{figure}[htbp]
  \begin{center}
    \begin{tabular}{c}

      \begin{minipage}{0.5\hsize}
        \begin{center}
          \includegraphics[scale=0.75]{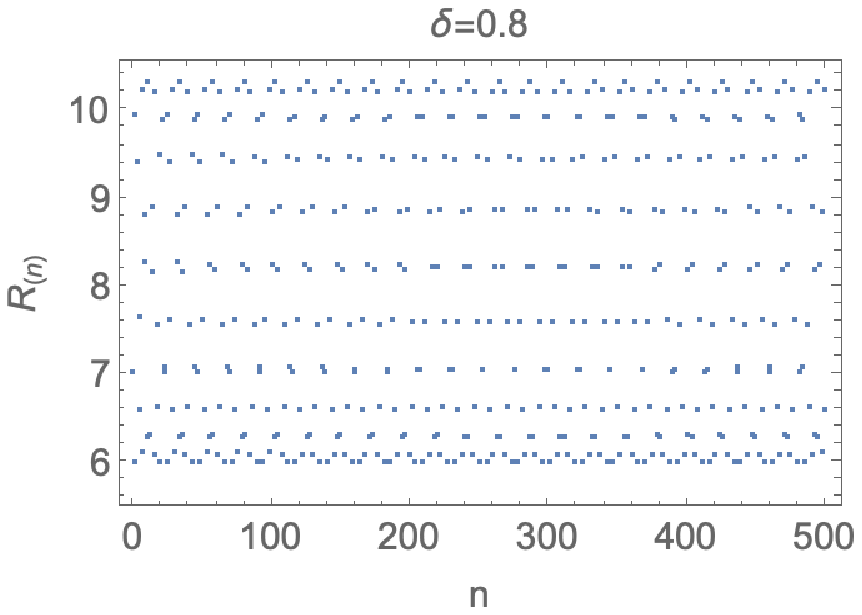}
         
        \end{center}
      \end{minipage}
      
      \begin{minipage}{0.5\hsize}
        \begin{center}
          \includegraphics[scale=0.75]{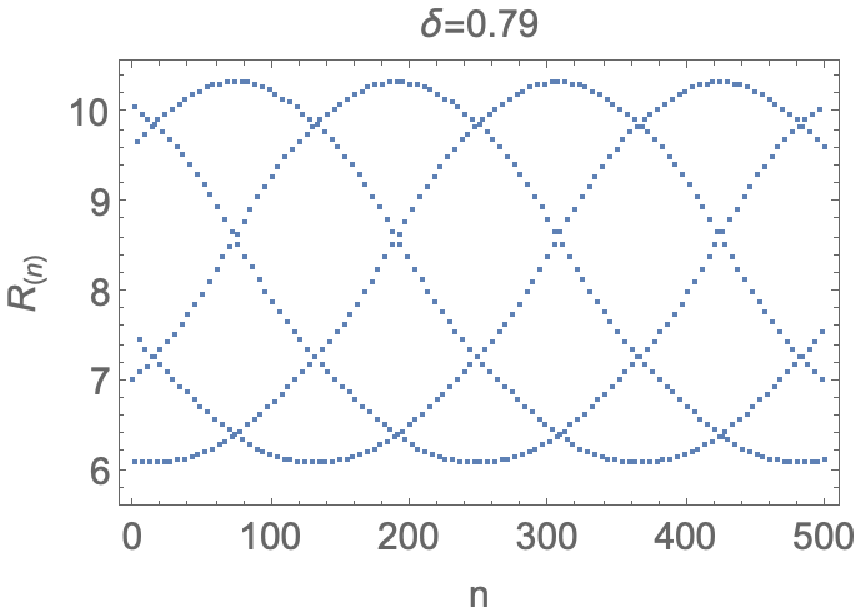}
         
        \end{center}
      \end{minipage}
 
\\

      \begin{minipage}{0.5\hsize}
        \begin{center}
          \includegraphics[scale=0.75]{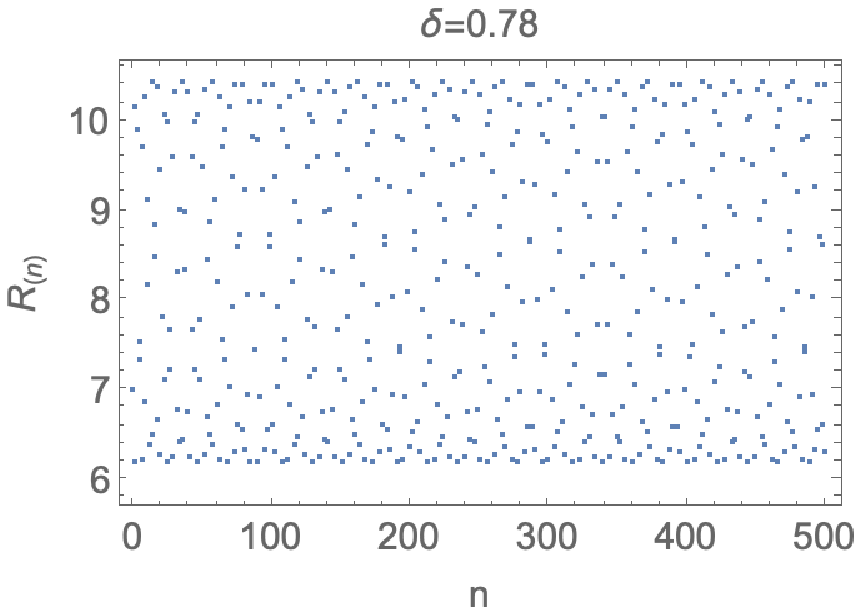}
          
        \end{center}
      \end{minipage}
      
      \begin{minipage}{0.5\hsize}
        \begin{center}
          \includegraphics[scale=0.75]{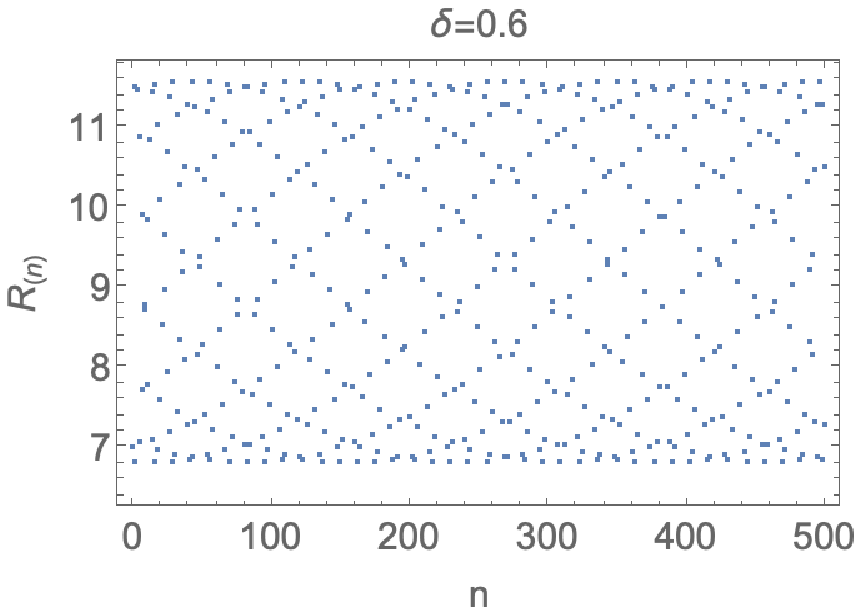}
         
        \end{center}
      \end{minipage}
 
\\
    \end{tabular}
    \caption{Crossing radius $R_{(n)}$ as a function of numbers of crossing $n$ for $\delta=0.8, 0.79, 0.78$ and $0.6$. These figures indicate sensitive behavior of $R_{(n)}$.}    
\label{fig-R}
  \end{center}
\end{figure}

When the shell oscillates stably, the evolution of $R_{(n)}$ varies considerably with the value of the initial energy $\delta$ and exhibits a rich behavior.
\fig{fig-R} shows crossing radius $R_{(n)}$ as a function of number of crossings $n$ for $\delta=0.8, 0.79, 0.78$ and $0.6$, with the help of the recursion relations of \eq{Rn} and \eq{DTR}. The first 500 crossings are shown.
These figures indicate sensitive behavior of $R_{(n)}$ ($M_{2(n)}$ shows qualitatively same behaviors as $R_{(n)}$, although not explicitly plotted in this paper). 
For certain values of $\delta$, $R_{(n)}$ (and also $M_{2(n)}$) may transit almost continuously. This means that the gravitational energy $\varepsilon$ of each shell has effectively continuous transitions (see $\delta=0.6, 0.79$ in \fig{fig-R}).
On the other hand, for another choice of $\delta$, $R_{(n)}$ behaves strangely, staying within a limited and very narrow range (see $\delta=0.8$ in \fig{fig-R}).
\begin{figure}[htbp]
  \begin{center}
          \includegraphics[scale=0.9]{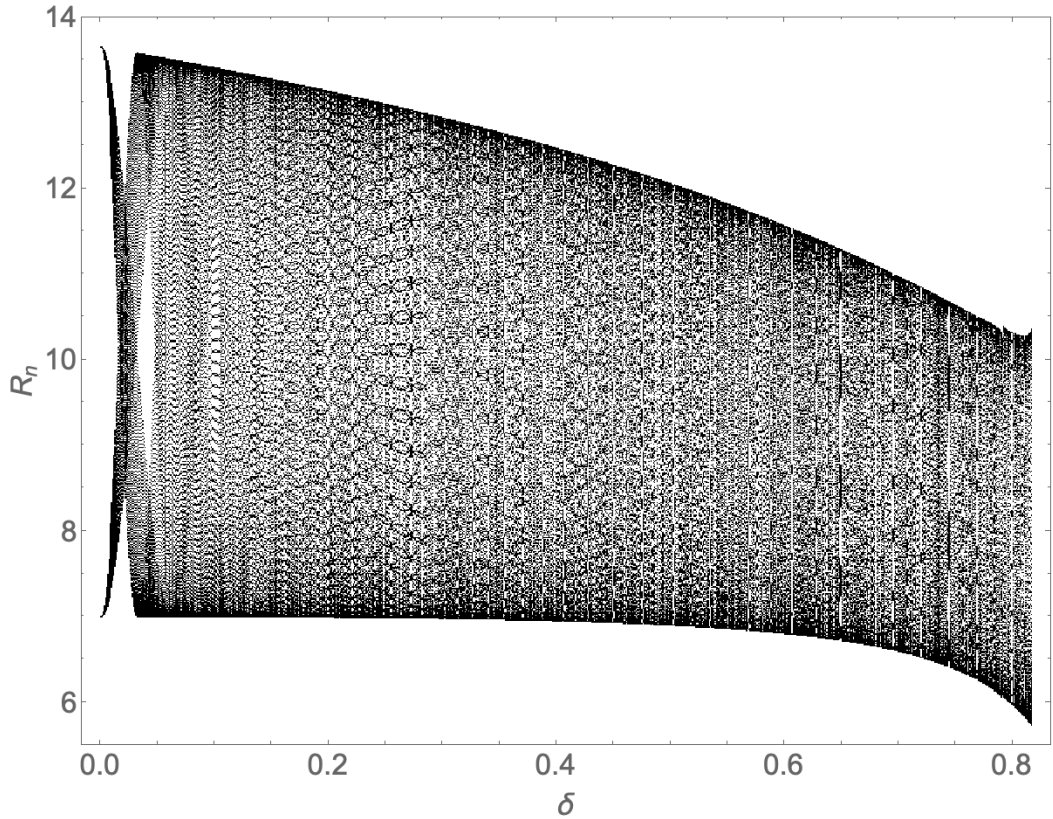}
    \caption{Crossing radii as a function of $\delta$ in stable periodic evolutions. Values of $\delta$ corresponding to stable motions spread into $0\leq \delta \lesssim 0.817$. Nontrivial stable motions correspond to $0.5 < \delta \lesssim 0.817$.
    \label{fig-bifurcationR7}}
  \end{center}
\end{figure}
Such a rich structure of crossing radii as a function of all stable $\delta$  is comprehensively seen in \fig{fig-bifurcationR7}. 
All plotted trajectories correspond to stable motions \cite{note2}.

\begin{figure}[htbp]
  \begin{center}
    \begin{tabular}{c}

      \begin{minipage}{0.5\hsize}
        \begin{center}
          \includegraphics[scale=0.6]{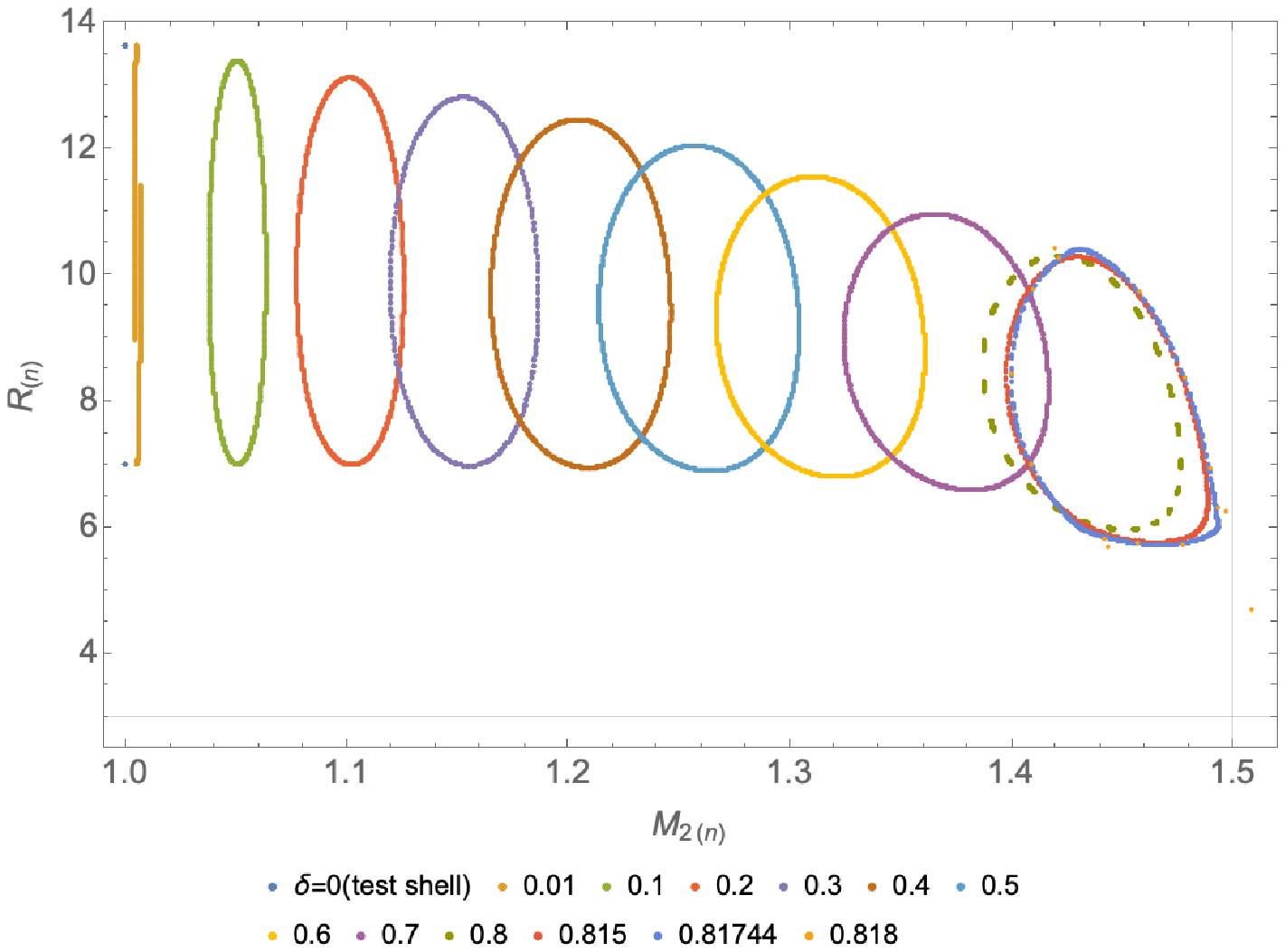}
         
        \end{center}
      \end{minipage}
      
      \begin{minipage}{0.5\hsize}
        \begin{center}
          \includegraphics[scale=0.5]{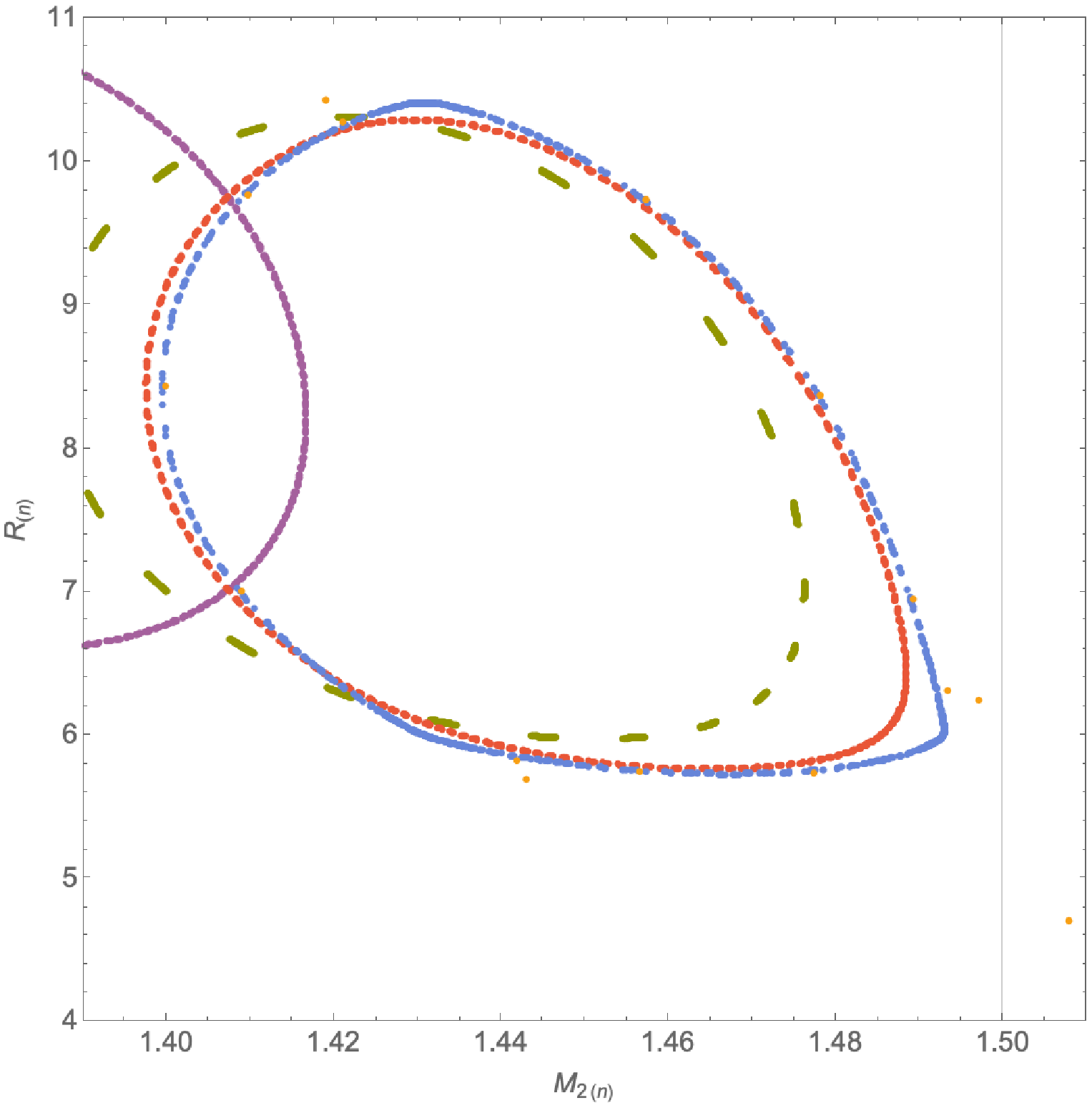}
         
        \end{center}
      \end{minipage}
 
\\
    \end{tabular}
    \caption{{\rm Left panel:} $R_{(n)}-M_{2(n)}$ plane for each $\delta$. A BH forms when the the point $(R_{(n)}, M_{2(n)})$ at $n$-th crossing crosses either the horizontal ($R_{(n)}=3$) or vertical straight line ($M_{2(n)}=1.5$).    {\rm Right panel:} zoom-in picture of the left panel.}    
\label{fig-MR}
  \end{center}
\end{figure}

As a final analysis, let us see how the trajectory $(R_{(n)}, M_{2(n)})$ of the recurrence equations changes from stable periodic motion to BH formation in the $R_{(n)}-M_{2(n)}$ plane.
\fig{fig-MR} represents the trajectory of the recurrence equations (\eq{Rn} and \eq{DTR}) for each delta.
 The horizontal straight line is the gravitational radius of the outer shell, $R_{(n)}=2M_3$. The vertical straight line is $M_{2(n)}=r_{in}/2$. When either $M_{2(n)}$ or $R_{(n)}$ crosses these lines, a BH forms (\eq{BH-criterion}).
When $\delta=0$, it describes the motion of the test shell, and in this case the crossing points are restricted to two specific radii determined by the initial values. See Appendix for the exact solution of the test shell's recurrence equations.
For larger $\delta$, the orbit is circular. As $\delta$ is further increased, the circular orbit approaches the two straight lines of the BH criterion. At the threshold value ($\delta=0.81744$), the circular orbit is disrupted and a cusp is formed in a part of the orbit (see the right panel in \fig{fig-MR} for $\delta=0.81744$).
At $\delta$ above the threshold, for a finite $n$, the orbit meets the BH formation criterion and exceeds the lines. This signals BH formation.

\newpage
\section{Summary and Discussion}\label{sec-conclusion}
The motion of two timelike shells confined between boundaries in strong gravity was numerically investigated.
Non-trivial stable motions of the shells were found with initial values for which a BH could form.
It was explicitly shown that stable motions are more likely to occur when the shells move at a high speed close the speed of light. 

In stable motions, by energy transfer, energy is accumulated in one shell at first, but after a while, the transfer in the opposite direction begins, and the energy begins to accumulate in the other shell. This energy transfer and reverse transfer are alternately repeated. As a result, neither shell becomes energetic enough to be BH.
On the other hand, at the initial value which eventually becomes BH, energy gradually accumulates in one shell, eventually promoting BH formation.

To a large extent, the quasi-periodic motion appears to be almost pure periodic motion. Considering the fact that two-shell systems are generally associated with chaotic properties \cite{MillerYoungkins1997, Barkov+2005, Kokubu:2020jvd}, the existence of such non-chaotic oscillations is somewhat surprising.
Although two-shell systems generally exhibit chaotic behavior even in Newtonian, our study shows that chaotic nature behind gravity disappears when the speed of the shell is quite relativistic.

The motion of a null shell as the null limit of a timelike shell is investigated using ``exact'' recurrence equations. 
Even for null shells as the fastest limit of timelike shells, the relation between initial conditions and stable motion was found to be non-trivial, presenting us with a rich behaviors.

As we have seen, the recurrence equations exhibit a rich behavior depending on the initial parameters, so it is difficult to solve the general term. Since the general term is not known, it seems impossible to determine anlitically the end state of the motion (BH or stable) from the initial data. However, since the obtained recurrence equations is exact, it is possible to investigate the end state of the system algorithmically by applying the equation to the given initial values in sequence with arbitrary precision.

\begin{figure}[htbp]
  \begin{center}
          \includegraphics[scale=0.6]{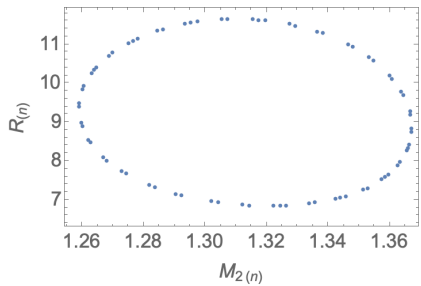}
    \caption{Development of $(R_{(n)},M_{2(n)})$ of timelike shells with $\delta=0.6$ in Type 2, forming a circular orbit as in the null shells.
              \label{fig-timelikeM2R}}
  \end{center}
\end{figure}

All of the stable quasi-periodic motions of timelike shells found in this study were found only when the shells were quite fast. Needless to say, a fast timelike shell can be regarded as almost a null shell, so we presume that the periodic motion of timelike shells is {\it essentially the same} as the stable periodic motion of null shells. In fact, as an example, the $R_{(n)}-M_{2(n)}$ plane orbit of the timelike shell is \fig{fig-timelikeM2R}, which has the same characteristics of a circular orbit as that of the null shell (\fig{fig-MR}).

Finally, we discuss the validity of the boundaries. In our setup, we have placed a rigid inner and outer boundary, which is somewhat artificial. A more physical model that provides an outer wall is to consider an asymptotically anti de-Sitter spacetime. In this spacetime, infinity can be reached in finite time, providing a natural outer wall. On the other hand, there are several physical factors that form an inner wall. A charged shell moving around a charged central body is repelled from the center due to an inner potential barrier. 
Matter with angular momentum can also form an inner barrier as well. Shells composed of collisionless particles with angular momentum effectively form an inner barrier. 
The BH formation/stable motion of two confined shells in these more physical situations is beyond the scope of this paper.

\acknowledgments
The author is grateful to C. Yoo, Y. Koga and T. Harada for fruitful discussions on the early stage of the investigation.
This work was supported by JSPS KAKENHI Grants No. JP20H05853 from the Japan Society for the Promotion of Science and the NSFC under Grants No. 12005059.

\appendix
\section{Test shell limit of the recurrence equations}
The simultaneous recurrence equation of confined null shells is given by \eq{Rn}, \eq{DTR}. 
It seems difficult to find the general term of this relation, but under the test shell limit, the general term can be exactly obtained.
The test null shell limit is obtained by setting $M_{2(n)}=M_1=M_3$. In this case, $M_2$ becomes a constant, and the DTR relation \eq{DTR} becomes trivial. The remaining equation \eq{Rn} becomes
\begin{align}
R_{(n+1)}=2M_1\left(1+W\left(\chi_n \right)\right),
\quad \chi_n:= c_1\left(\frac{R_{(n)}}{2M_1}-1\right)^{-1} \exp\left(c_2-\frac{R_{(n)}}{2M_1}\right),
\label{test-Rn}
\end{align}
where $c_{1,2}$ is constants and represents the $R_{(n)}$-independent term of the argument of $W$ in \eq{Rn}. Introducing $a_n=R_{(n)}/(2M_1)-1$, \eq{test-Rn} reduces to
\begin{align}
a_{n+1}=W(\frac{c_3}{a_ne^{a_n}}),
\end{align}
where $c_3=c_1e^{c_2}$ is constant. The inverse solution of the above equation is
\begin{align}
\frac{c_3}{a_ne^{a_n}}=W^{-1}(a_{n+1})=a_{n+1}e^{a_{n+1}}.
\end{align}
The last equality used $W^{-1}(x)=xe^x$.
If we let $b_n=a_n e^{a_n}$, we arrive
\begin{align}
b_{n+1}=\frac{c_3}{b_n}. \label{test-recursion}
\end{align}
\eq{test-recursion} generally represents a simple oscillating solution with the two values $b_0, b_1, b_0,b_1, \cdots$. A special case is when the first term is $b_0=\sqrt{c_3}$, resulting $b_n=\sqrt{c_3}$. This is an identity map, and the shells intersect at the same radius each time.

From the above, we proved that test null shells generally intersect at two specific radii.



\begin{thebibliography}{99}
\bibitem{LIGOScientific:2016aoc}
B.~P.~Abbott \textit{et al.} [LIGO Scientific and Virgo],
Phys. Rev. Lett. \textbf{116}, no.6, 061102 (2016)

\bibitem{Oppenheimer-Snyder}
J.~R.~Oppenheimer and H.~Snyder,
  Phys.\ Rev.\  {\bf 56}, 455 (1939).

\bibitem{Bizon:2011gg}
P.~Bizon and A.~Rostworowski,
Phys. Rev. Lett. \textbf{107}, 031102 (2011) . 

\bibitem{Buchel:2012uh}
A.~Buchel, L.~Lehner and S.~L.~Liebling,
Phys. Rev. D \textbf{86}, 123011 (2012).

\bibitem{Maliborski:2012gx}
M.~Maliborski,
Phys. Rev. Lett. \textbf{109}, 221101 (2012).

\bibitem{Maliborski:2013jca}
M.~Maliborski and A.~Rostworowski,
Phys. Rev. Lett. \textbf{111}, 051102 (2013).

\bibitem{Okawa:2014nea}
H.~Okawa, V.~Cardoso and P.~Pani,
Phys. Rev. D \textbf{90}, no.10, 104032 (2014).

\bibitem{Okawa:2015xma}
H.~Okawa, J.~C.~Lopes and V.~Cardoso,
[arXiv:1504.05203 [gr-qc]].

\bibitem{Israel1966}
W.~Israel,
Nuovo Cim. B \textbf{44S10}, 1 (1966).

\bibitem{DrayHooft1985}
T.~Dray and G.~'t Hooft,
Commun. Math. Phys. \textbf{99}, 613-625 (1985).

\bibitem{redmount}
I. H. Redmount, Progress of Theoretical Physics {\bf 73}, 1401 (1985).

\bibitem{PoissonIsrael1989}
E.~Poisson and W.~Israel,
Phys. Rev. Lett. \textbf{63}, 1663-1666 (1989).

\bibitem{Nunez+1993}
D.~Nunez, H.~P.~de Oliveira and J.~Salim,
Class. Quant. Grav. \textbf{10}, 1117-1126 (1993).

\bibitem{NakaoIdaSugiura1999}
K.~i.~Nakao, D.~Ida and N.~Sugiura,
Prog. Theor. Phys. \textbf{101}, no.1, 47-71 (1999).

\bibitem{IdaNakao1999}
D.~Ida and K.~i.~Nakao,
Prog. Theor. Phys. \textbf{101}, no.5, 989-1000 (1999).

\bibitem{Barkov+2002JTEP}
M.~V.~Barkov, V.~A.~Belinski and G.~S.~Bisnovatyi-Kogan,
J. Exp. Theor. Phys. \textbf{95}, 371-391 (2002).

\bibitem{CardosoRocha2016}
V.~Cardoso and J.~V.~Rocha,
Phys. Rev. D \textbf{93}, no.8, 084034 (2016).

\bibitem{BritoCardosoRocha2016}
R.~Brito, V.~Cardoso and J.~V.~Rocha,
Phys. Rev. D \textbf{94}, no.2, 024003 (2016).

\bibitem{MillerYoungkins1997}
B. N. Miller and V. Youngkins, 
Chaos \textbf{7}, 187 (1997).

\bibitem{Barkov+2005}
M. V. Barkov, G. S. Bisnovatyi-Kogan, A. I. Neishtadt, and V. A. Belinski, 
Chaos \textbf{15}, 013104 (2005).

\bibitem{Kokubu:2020jvd}
T.~Kokubu,
Phys. Rev. D \textbf{102}, no.6, 064032 (2020).

\bibitem{Mazur:2001fv}
P.~O.~Mazur and E.~Mottola,
Universe \textbf{9}, 88 (2023).

\bibitem{Visser:2003ge}
M.~Visser and D.~L.~Wiltshire,
Class. Quant. Grav. \textbf{21}, 1135-1152 (2004).

\bibitem{Pfister:1991ky}
H.~Pfister,
Class. Quant. Grav. \textbf{13}, 2267-2277 (1996).

\bibitem{Pani:2009ss}
P.~Pani, E.~Berti, V.~Cardoso, Y.~Chen and R.~Norte,
Phys. Rev. D \textbf{80}, 124047 (2009).

\bibitem{note1}
Carefully looking at \fig{fig-type1}, we realize there are a few stable solutions in the range where $\delta$ is a little larger than $0.5$. However, it is questionable whether these particular solutions really represent stable solutions. The reasons are as follows. 
Evolution of timelike shells with small $\delta$ indeed shows typical chaotic behavior (for large $\delta$, we do not see any chaos, but periodic motion). If the motion is chaotic, then BHs always form if a sufficiently long integration time is taken. 
Therefore, we can assume that these solutions will probably be BH if we take a long enough integration time. Our numerical calculations only show that these solutions are stable up to $t=2000$.

\bibitem{note2}
Mathematically, this figure is a so-called bifurcation diagram. If the area taken by the vertical axis $R_{(n)}$ suddenly expands as the parameter $\delta$ changes (like a well-known logistic map), we may conclude that this system is chaotic. However, in the present case, the system does not show such a sudden increase in the vertical axis (although it shows complicated patterns), so we can say that this system is not chaotic.
This is reasonable because we consider null shells, not timelike shells. In (slow) timelike shells, chaos generally  appears (see Ref.\cite{note1}).

\bibitem{Corless:1996zz}
R.~M.~Corless, G.~H.~Gonnet, D.~E.~G.~Hare, D.~J.~Jeffrey and D.~E.~Knuth,
Adv. Comput. Math. \textbf{5}, 329-359 (1996).
\end{thebibliography}
\end{document}